\documentclass[%
reprint,
amsmath,amssymb,
prab,
floatfix,
]{revtex4-2}

\usepackage[USenglish]{babel}

\usepackage{graphicx}
\usepackage{xcolor} 
\usepackage{soul} 
\usepackage{siunitx} 
\usepackage{bigints} 

\begin{document}

\title{Corrections for systematic errors in slit-profiler transverse phase space measurements}

\author{C. Richard}
	\email{Contact author: christopher.richard@desy.de}
	\affiliation{Deutsches Elektronen Synchrotron DESY, Platanenallee 6, 15738 Zeuthen, Germany}
\author{M. Krasilnikov, N. Aftab, Z. Amirkhanyan, D. Dmytriiev, A. Hoffmann, X.-K. Li, Z. Lotfi, F. Stephan, G. Vashchenko, S. Zeeshan}
	\affiliation{Deutsches Elektronen Synchrotron DESY, Platanenallee 6, 15738 Zeuthen, Germany}

\begin{abstract}
	In photo injectors, the transverse emittance is one of the key measures of beam quality as it defines the possible performance of the whole facility. As such it is important to measure the emittance in photo injectors and ensure the accuracy of these measurements. While there are many different methods of measuring the emittance, this paper focuses on quantifying the systematic errors present in transverse phase space measurements taken with slit-profiler methods, i.e. scanning a narrow slit over the beam and continually measuring the passed beamlets' divergence with a downstream profiler. The measurement errors include effects of the slit size, beamlet imaging, and residual space charge. While these effects are generally small, they can have significant impact on the measured emittance when the 2D phase space is strongly coupled. The systematic effects studied and corrections are demonstrated with simulations and measurements from the Photo Injector Test facility at DESY in Zeuthen (PITZ) using a slit-screen emittance scanner.
\end{abstract}

\date{\today}

\maketitle

\section{Introduction}
Photo injectors are electron sources capable of delivering short bunches with high brightness \cite{Photoinjectors}. This makes them ideal for use in x-ray free electron lasers (XFELs) \cite{XFELs}, ultrafast electron diffraction \cite{Electron_diffraction}, energy recovery linacs \cite{ERLs}, and electron cooling \cite{electron_cooling}. 

One of the key parameters of electron beam quality in photo injectors is the transverse emittance $\epsilon$ which characterizes the beam area in phase space \cite{Wiedemann}. It is necessary to maintain sufficiently small emittance for beamline matching, reducing losses, and increasing beam brightness. In photo injectors, which are linear accelerators, the emittance can only increase along the accelerator. Generally, the emittance at the end of the photo injector is of most importance as this is where the beam will be injected into a larger accelerator or used for experiments. However, measuring emittance near the source is also crucial as this sets the minimum possible emittance before it increases throughout the accelerator due to space charge, collective effects, bunch compression, and non-linear applied fields. 

\begin{figure}
	\centering
	\includegraphics[width=3.25in]{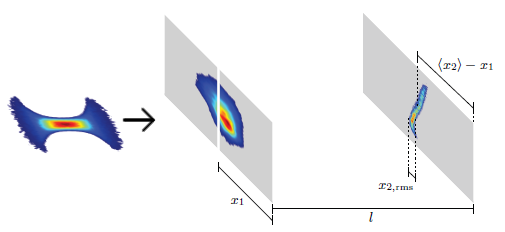}
	\caption{Slit-screen phase space measurement scheme. The slit allows only a small beamlet to pass through which is measured downstream. Shown here is a vertical slit which is used to measure the horizontal phase space \cite{Raffael_thesis}.}
	\label{fig:slit_scanner}
\end{figure}

There are many methods for measuring the emittance including quadrople or solenoid scans \cite{Quad_scan}, tomography \cite{Tomography}, and slit-and-profiler methods. At low energy, near the source, slit-and-profiler methods are preferred because they significantly reduce the effect of space charge compared to methods such as quadrupole scans \cite{Emittance_space_charge}. There are multiple different types of slit-and-profiler methods including pepper pots, Allison scanners \cite{PIP2_allison_scanner}, slit-slit, slit-harp \cite{Slit_harp}, and slit-screen \cite{Slit_scan}. They all follow the same basic principle of reconstructing an image of the transverse phase space by measuring the local divergence as a function of position then combining these measurements to construct the beam's transverse, 2D phase space. This is done by inserting a thin slit into the beam to only allow particles at a specific transverse location to pass (Fig. \ref{fig:slit_scanner}). The passed beamlet is then measured with a profiler far enough downstream such that the beamlet divergence dominates the position profile and the local divergence of the beam at the slit position $I(u'_\text{slit})$ is given by 
\begin{equation}
	I(u'_\text{slit}) = I(u_\text{profiler})/L,
	\label{eqn:pos_to_angle}
\end{equation}
where  $u$ denotes either the $x$ or $y$ plane, $I(u'_\text{slit})$ is the location divergence at the slit, $I(u_\text{profiler})$ is the intensity profile at the profiler and $L$ is the distance between the slit and the screen. The slit is stepped across the beam and the measured $I(u_\text{profiler})$ at each location are combined to construct the phase space image. Note that determining the correct central divergence at each slit requires accounting for the position of the slit. This is done by shifting the measured profile by the slit position to center all the profiles. From this imaging, the geometric emittance can be calculated as
\begin{equation}
	\epsilon_g = \sqrt{\sigma_u^2 \sigma_{u'}^2 - \sigma_{uu'}^2}
\end{equation}
where $\sigma_u$ is the rms beam size, $\sigma_{u'}$ is the rms divergence, and $\sigma_{uu'}$ is the $\left<uu'\right>$ covariance. Typically, the geometric emittance is not cited, but rather the normalized emittance $\epsilon_n$ to compare beams with different energies. The normalized emittance is
\begin{equation}
	\epsilon_n = \beta \gamma \epsilon_g
\end{equation}
where $\beta$ is the normalized velocity and $\gamma$ is the Lorentz factor. 

Accurate measurements of the emittance from these emittance scanners are necessary to properly characterize the performance of photoinjectors. Therefore, detailed studies of the systematic errors in the scanners are necessary. Several studies of the systematic errors in the slit-screen scanners have been performed \cite{Staykov_thesis, Spesyvtsev_thesis, Qian_slit_scan}. It was found that the primary systematic errors in this system are the resolution and dynamic range of the camera and residual space charge in the beamlets. However, these studies do not propose corrections for the systematic errors and it is desired to correct for these effects to accurately determine the transverse emittance. In addition, these studies focused on the effects of the camera resolution on profile measurements of the full beam, not beamlet measurements, and do not quantify the effects on the measured emittance. The effect of the slit width on the measured emittance has also been estimated \cite{Raffael_thesis, Grygorii_thesis}, but, for simplicity, it was assumed the beam had no {$u-u'$} coupling and the camera effects were either excluded or simplified. More complete corrections for the slit width have been studied for Allison scanners \cite{Wong_AS_correction, Fermilab_slit_corr}, but they are not directly applicable to single slit methods.

In this paper the systematic errors in the slit-screen emittance are studied starting with the effect of the slit, then effects of imaging, and lastly space charge and noise cuts. The focus is on characterizing and correcting the errors in an existing system rather than optimizing for a new design. Therefore, topics such as choosing an appropriate drift length, scintillator material \cite{PITZ_nonlinear_screen1, PITZ_nonlinear_screen2}, and adjusting the slit thickness for uniform noise on the screen \cite{PITZ_EMSY_design} are not discussed. In addition, errors introduced by variations in the incoming beam, e.g. position jitter \cite{Jitter_emittance} and variations along the bunch train, are not taken into account.

\subsection{Slit-screen measurements at PITZ}
The Photo Injector Test facility at DESY in Zeuthen (PITZ) serves as a dedicated research and development platform for RF photo-injectors, the related technologies, and applications. It has successfully demonstrated the electron source requirements for the Free-electron LASer in Hamburg (FLASH) and the European XFEL \cite{PITZ_overview}. Since the start of operations in 2002, the PITZ beamline has been continuously evolving to meet the research demands for the European XFEL, focusing on three primary objectives: developing, conditioning, and characterizing RF guns for both user facilities.

The transverse emittance is one of the crucial beam properties in photo-injectors. A lower emittance improves the performance of X-ray free-electron lasers in terms of highest photon flux or achievable shortest wavelength. As such, the transverse emittance is a crucial parameter measured at PITZ to judge the brightness of the electron bunch and allows for direct comparison of simulation and experimental measurements. In addition, PITZ has studied photo cathode laser pulse shaping in simulations and experiments as a method for reducing the transverse emittance \cite{laser_shaping}.

\begin{figure}
	\centering
	\includegraphics[width=\columnwidth]{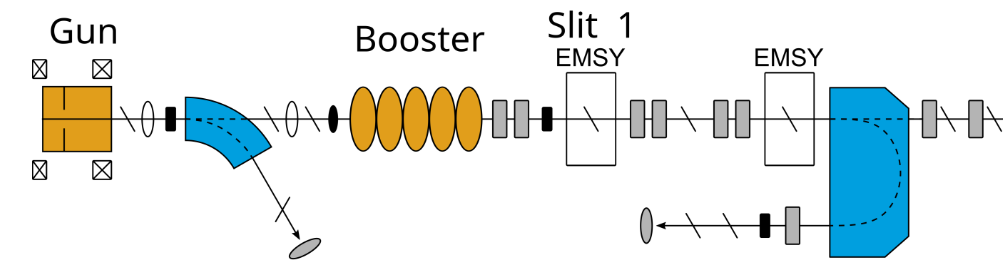}
	\caption{Section of PITZ beamline used for emittance measurements. The gun is surrounded by a main and bucking solenoid for focusing. Two dipoles are used for energy measurements before and after the booster. The diagonal lines represent screen stations and the screen directly after the second dipole is used for imaging the beamlets. The slits for emittance measurements are inserted at `EMSY' (Emittance Measurement SYstem).}
	\label{fig:PITZ}
\end{figure}

Emittance measurements at PITZ are taken near the exit of the RF booster at 17-\SI{20}{MeV/c} using a \SI{50}{\micro m} slit to create the beamlets and a YAG screen located \SI{3.133}{m} downstream of the slit to measure the beamlet profiles (Fig. \ref{fig:PITZ}). In addition to the beamlet images, for every measurement an image of the whole beam is taken at the slit position using a YAG screen, this image is referred to here as the spot image. 

To find the minimum emittance for a given bunch charge and photocathode laser pulse shape and length \cite{PITZ_laser_shaping}, sets of emittance measurements are taken with different photogun solenoid strengths and laser spot sizes on the cathode. This procedure finds the optimal space charge density at the cathode and optimizes the optics of the emittance compensation scheme to minimize the projected transverse emittance after the booster \cite{Emittance_compensation}. After the booster, the higher energy significantly reduces the space charge effects resulting in a more stable emittance throughout the rest of the beamline. This paper uses emittance measurements of gun solenoid scans at PITZ as examples because they cover a range of emittances and Twiss parameters and these scans are a representative model of a standard measurement.

\section{Effect of the slit}
\subsection{Slit width}
Slit-based emittance measurements rely on having a sufficiently narrow slit to convert a space charge dominated beam into a divergence dominated beamlet.  With a single slit scan procedure, the slit opening $\Delta_s$ must be large enough to have sufficient angular acceptance to measure the full phase space and also allow enough particles to pass through to ensure sufficient signal. However, the slit cannot be too wide because the measured second-order moments depend on the slit size.

The effect of the slit on the measurement can be calculated for a Gaussian beam phase space 
\begin{align}
	P(u,u')& = \nonumber \\
	&\frac{1}{2\pi\epsilon_g}\exp\left[ -\frac{1}{2\epsilon_g^2}\left(\sigma_{u'}^2 u^2 - 2\sigma_{uu'}uu' + \sigma_u^2 u'^2 \right)\right].
\end{align}
For simplicity, the coordinates are normalized
\begin{equation}
	\bar{u} = \frac{\sigma_{u'}}{\epsilon_g} u,~~\bar{u}' = \frac{\sigma_u}{\epsilon_g}u', ~~ \nu=\frac{\sigma_{uu'}}{\sigma_u\sigma_{u'}}
\end{equation}
where $\nu$ is the relative strength of the coupling compared to the rms beam size and divergence. Note that $|\nu|\leq1$ with $\nu=0$ corresponds to no $u-u'$ coupling. The beam distribution becomes
\begin{equation}
	P(\bar{u},\bar{u}') = \frac{1}{2\pi\epsilon_g}\exp\left[ -\frac{1}{2} \left(\bar{u}^2 - 2\nu\bar{u}\bar{u}' + \bar{u}'^2 \right)\right].
\end{equation}
The particles' positions can be propagated to the profiler
\begin{equation}
	u_{p} = u + u' L \label{eqn:linear_prop}
\end{equation}
\begin{equation}
	\bar{u}_{p} \equiv u_{p} \frac{\sigma_{u'}}{\epsilon_g} = \bar{u} + \frac{\bar{u}'}{\mu}
	\label{eqn:beamlet_coord_norm}
\end{equation}
where subscript $p$ denotes coordinates at the profiler and $\mu$ is the ratio of the contributions of initial beam size and the divergence contribution to the beam size at the profiler
\begin{equation}
	\mu = \frac{\sigma_u}{\sigma_{u'}L}.
\end{equation}
The beam distribution at the profiler then becomes
\begin{multline}
	P(\bar{u},\bar{u}_{p}) =\\
	\frac{1}{2\pi\epsilon_g}\exp\left[-\frac{1}{2}\left(\bar{u}^2 - 2\nu\mu \bar{u}(\bar{u}_{p}-\bar{u}) + \mu^2(\bar{u}_{p}-\bar{u})^2 \right)\right].
\end{multline}
This distribution is integrated over the slit opening to get the beamlet distribution at the profiler $\rho_{bl,p}$ for a slit centered at $\bar{u}_0$
\begin{align}
	\rho_{bl,p}&(\bar{u}_0,\bar{u}_{p}) = \int_{\bar{u}_0-\delta/2}^{\bar{u}_0+\delta/2} \text{d}\bar{u}~ \frac{\epsilon}{\sigma_{u'}} P(\bar{u},\bar{u}_{p}) \\
	\rho_{bl,p}&(\bar{u}_0,\bar{u}_{p}) = \nonumber \\ 
	&\exp\left[\frac{\mu^2 \bar{u}_{p}^2 (\nu^2-1)}{2(1+2\nu\mu+\mu^2)}\right] \frac{\left[ \text{erf}(\Lambda_+)-\text{erf}(\Lambda_-) \right]} {2\sigma_{u'}\sqrt{2\pi}\sqrt{1+2\nu\mu+\mu^2}} 
	\label{eqn:beamlet_dist_norm}
\end{align}
where $\delta$ is the normalized slit opening
\begin{equation}
	\delta = \frac{\sigma_{u'}}{\epsilon_g}\Delta_s
\end{equation}
and 
\begin{equation}
	\Lambda_\pm = \frac{(\bar{u}_0 \pm \delta/2)(1+2\nu\mu+\mu^2)-\bar{u}_{p}\mu(\nu+\mu)} {\sqrt{2}\sqrt{1+2\nu\mu+\mu^2}}.
\end{equation}
This can be expanded to third order in $\delta$ then integrated over $\bar{u}_{p}$ to get the measured beamlet size $\sigma_{bl,m}$
\begin{equation}
	\left< \bar{u}_{bl,p}^2 \right> = \frac{1}{\mu^2} + \frac{(\nu + \mu)^2}{12\mu^2}\delta^2
\end{equation}
\begin{equation}
	\sigma_{bl,m}^2 = \left< \bar{u}_{bl,p}^2 \right>\frac{\epsilon_g^2}{\sigma_{u'}^2} =  \frac{\epsilon_g^2}{\sigma_u^2}L^2 + \frac{\Delta_s^2}{12}\left( 1+\frac{L\sigma_{uu'}}{\sigma_u^2} \right)^2. 
	\label{eqn:beamlet_size}
\end{equation}
Note that $\Delta_s/\sqrt{12}$ is the rms size of the slit opening. The measured phase space $\rho_{bl}(\bar{u}_0,\bar{u}')$ can now be reconstructed by substituting Eq. \ref{eqn:beamlet_coord_norm} into Eq. \ref{eqn:beamlet_dist_norm} to convert the positions on the profiler back to divergence. The measured second order moments, up to second order in $\delta$ can then be calculated
\begin{gather}
	\left<\bar{u}^2\right> =  \frac{1}{1-\nu^2} + \frac{\delta^2}{12} \\
	\left<\bar{u}'^2\right> = \frac{1}{1-\nu^2} + \mu^2\frac{\delta^2}{12} \\
	\left<\bar{u}\bar{u}'\right> = \frac{\nu}{1-\nu^2} - \mu\frac{\delta^2}{12}.
\end{gather}
Converting back to physical coordinates gives measured second order moments, denoted by subscript $m$
\begin{gather}
	\sigma_{u,m}^2 = \sigma_u^2+\frac{\Delta_s^2}{12} \label{eqn:slit_profile_effect} \\
	\sigma_{u',m}^2 = \sigma_{u'}^2 + \frac{\Delta_s^2}{12L^2} \label{eqn:slit_angle_effect}\\
	\sigma_{uu',m} = \sigma_{uu'} - \frac{\Delta_s^2}{12L} \label{eqn:slit_coupling_effect}.
\end{gather}

The effects of the slit on the measured beam size and divergence are, intuitively, convolutions of the true distributions with the slit opening. For bunch size measurements, this effect is generally negligible for a \SI{50}{\micro\m} slit and typical rms beam sizes at PITZ of $\gtrsim$\SI{0.2}{\mm} with $\mu\approx1.3$. The rms divergence at PITZ is $\gtrsim$\SI{0.05}{\milli rad}. In this case, with \SI{50}{\micro\m} slit and $L$=\SI{3.133}{m}, the slit will increase $\sigma_{u',m}$ by $\lesssim$\SI{5}{\%}.

\begin{figure}
	\centering
	\includegraphics{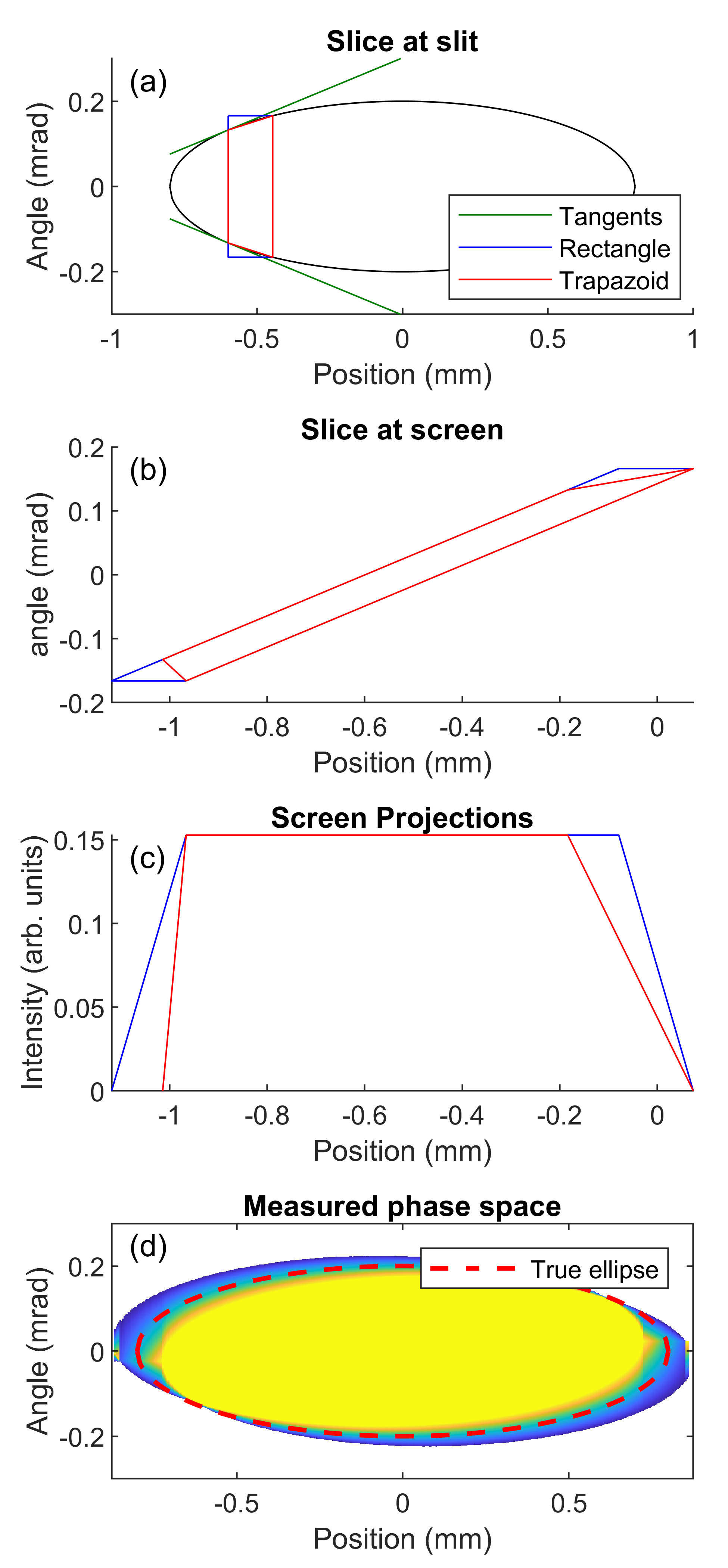}
	\caption{(a) The beamlet phase space can be modeled with a rectangle, however, a trapezoid more accurately models the slope of the phase space ellipse across the slit. (b) The resulting beamlet phase spaces at the profiler for a rectangular and trapezoidal beamlet. (c) The projection of the rectangular beamlet is symmetric, but the trapezoidal beamlet has an asymmetric projection that depends on the slopes of the edges. (d) The resulting reconstructed phase space using the trapezoidal beamlet for a uniform distribution shows a clear coupling due to the asymmetry caused by the slit opening.
	\label{fig:slit_coupling}}
\end{figure}

The effect of the slit on $\sigma_{uu',m}$ is caused by slight variations in intensity of the beam across the slit opening. For illustration, an exaggerated example of this effect is shown in Fig. \ref{fig:slit_coupling} for an uncoupled ($\nu=0$), uniform phase space distribution. Figure \ref{fig:slit_coupling}a shows two models of a beamlet. The first is a simple rectangular model and the second is a trapezoid with slopes of the top and bottom edges defined by the tangents of the phase space. These phase spaces are then propagated to the profiler (\ref{fig:slit_coupling}b). The rectangular slit results in a symmetric parallelogram, but the more realistic trapezoidal beamlet is asymmetric. This can be clearly seen in the position projections of the beamlet phase spaces at the profiler. This asymmetric effect varies across the beam as the slopes of the top and bottom of the phase space ellipse change. The net effect is an asymmetry being added to the measured phase space resulting in a systematic coupling effect.

The validity of Eqs. \ref{eqn:slit_profile_effect}-\ref{eqn:slit_coupling_effect} for realistic, non-Gaussian phase spaces can be studied by analyzing phase spaces simulated with ASTRA \cite{ASTRA} with a linear slit scanner model in Matlab \cite{matlab}. The slit scanner model replicates a slit-screen scan by selecting the particles at a given slit position then propagating them in $x$ and $y$ to the screen location using Eq. \ref{eqn:linear_prop}. A heat map is then generated from the resulting $x-y$ beamlet distribution to replicate profiling the beamlets with a screen. The bin size of the beamlet images is similar to what is used in beamline measurements, typically $\sim$\SI{40}{\micro m} at PITZ. This is repeated for all slit positions across the beam to replicate the beamlet images. These images can then be input into scripts for processing real measurements to reconstruct the phase space.

\begin{figure}
	\centering
	\includegraphics{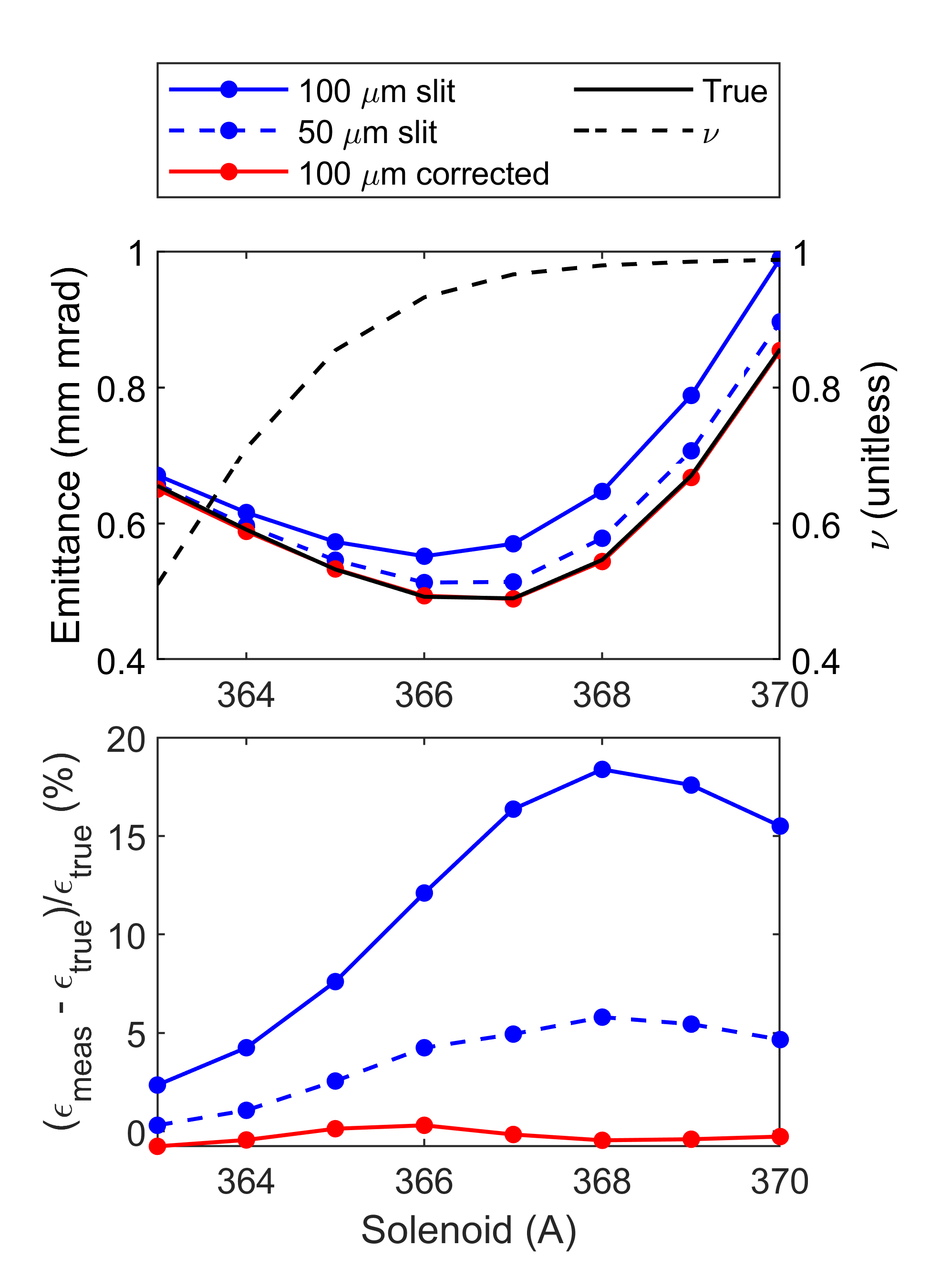}
	\caption{Simulated emittance using a \SI{50}{\micro m} and \SI{100}{\micro m} slit. After correction with Eqs. \ref{eqn:slit_profile_effect}-\ref{eqn:slit_coupling_effect} the measurements match the true emittance (\SI{50}{\micro m} corrected results not shown). \SI{50}{\micro m} slits are commonly used at PITZ for emittance measurements.}
	\label{fig:slit_correction}
\end{figure}

Figure \ref{fig:slit_correction} shows the result of ASTRA simulations of the PITZ beamline from the photo-cathode to the slit after the booster for \SI{250}{\pico C} beams with different gun solenoid strengths. The simulation results were processed with the linear slit scanner model with slit sizes of 50 and \SI{100}{\micro\meter}. Note that a \SI{100}{\micro\meter} slit is generally too large for measurements at PITZ, however, it is shown here to clearly demonstrate the effect. The rms second order moments of the two cases differ by $<1\%$ from the true values for all solenoid currents. However, the error from the true emittance in the \SI{50}{\micro\m} case is within 6\%  and the \SI{100}{\micro\meter} case increases to 20\% error. This increase occurs at higher solenoid current where the phase space is more strongly coupled, i.e. $\sigma_u\sigma_{u'/}\sigma_{uu'}\approx1$. In this case, the calculated emittance is the difference of two similar values, resulting in the increase sensitivity to each value. 

The slit stations at PITZ also have \SI{10}{\micro m} slits to reduce the effect of the slit opening. With \SI{10}{\micro m} slits, the increase in the measured emittance is reduced to $<2\%$. However, these narrower slits allow less of the beam to pass resulting in lower intensity beamlet images and more of the signal is lost under the noise floor. For this reason, the \SI{10}{\micro m} slits are not typically used.

However, the true second order moments can be determined using Eqs. \ref{eqn:slit_profile_effect}-\ref{eqn:slit_coupling_effect} to subtract off the effect of the slit.  After correction emittances from both slit sizes are within 1\% of the true values for all slit sizes. This allows for accurate measurements of the emittance with larger slits for improves SNR.

\subsection{Slit thickness}
\begin{figure}
	\centering
	\includegraphics{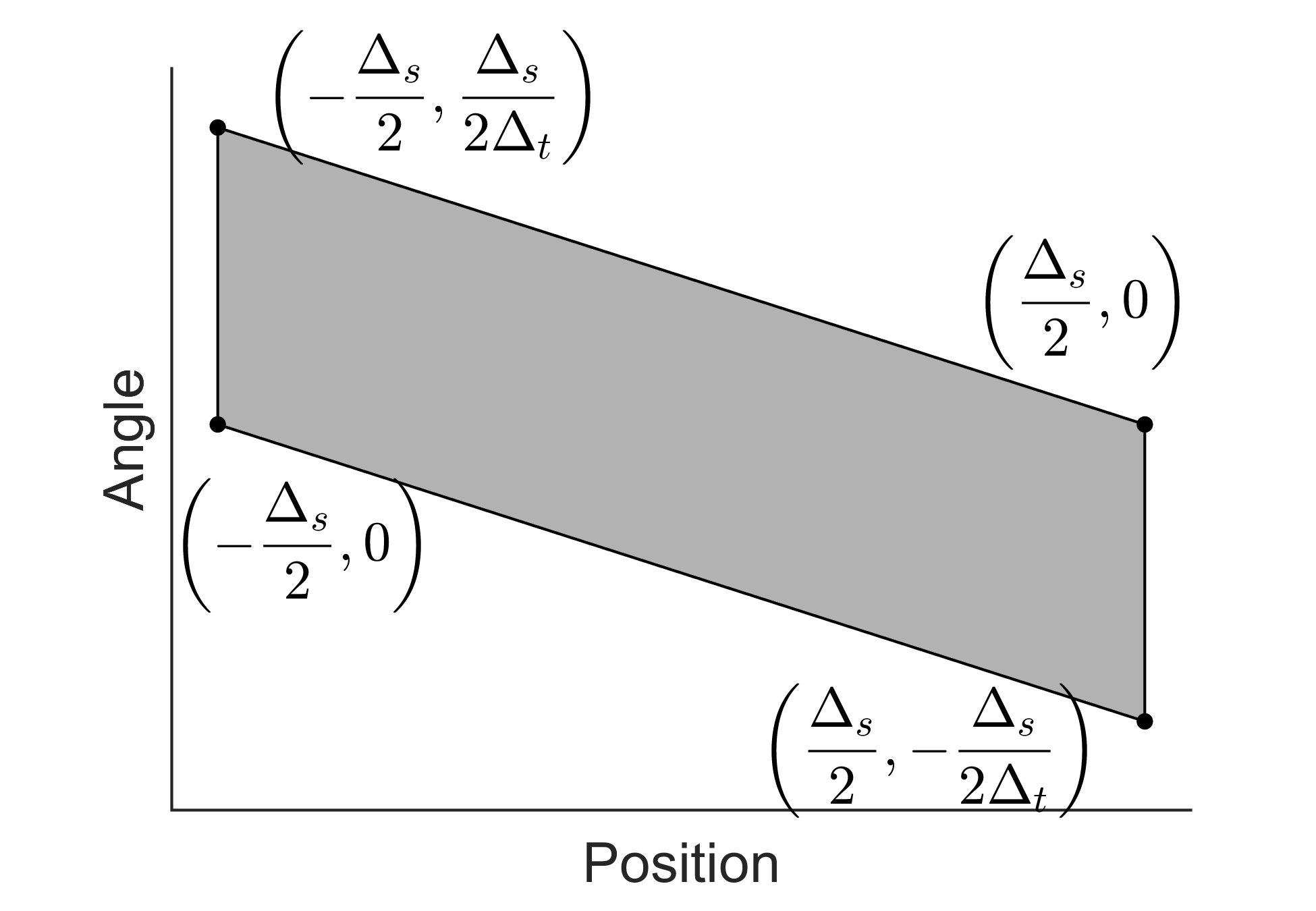}
	\includegraphics{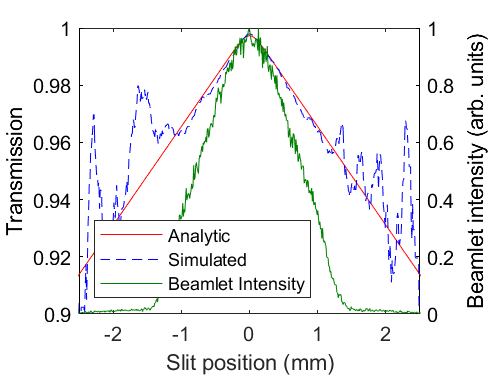}
	\caption{Top: Phase space acceptance of a slit in terms of the slit opening and thickness. The thickness can cut particles when the divergences are large. Bottom: example of simulated and analytic transmission of a \SI{1}{\nano C} beam with a \SI{10}{\micro m} wide, \SI{1}{\milli m} thick slit. The simulated results show a moving average to reduce noise.}
	\label{fig:slit_acceptance}
\end{figure}

The thickness of the slit $\Delta_t$ restricts the angular acceptance of the system. This results in a parallelogram phase space acceptance with vertices shown in Fig. \ref{fig:slit_acceptance} top similar to a two slit scanner \cite{Wong_AS_correction}. The angular acceptance of a single slit is large and therefore only removes particles with large divergences. This generally causes larger cuts on the edges of the beam where the central divergences of the beamlets are largest. The transmission through the slit can be estimated assuming, at the slit, the local angular distribution is Gaussian and position distribution is uniform 
\begin{equation}
	T = \frac{\bigintsss_{-\Delta_s/2}^{\Delta_s/2} \bigintsss_{(-\Delta_s/2-u)/\Delta_t}^{(\Delta_s/2-u)/\Delta_t} \text{d}u'\text{d}u~ e^\frac{-\left(u'-u'_{0,bl}\right)^2}{2\sigma_{u',bl}^2}}{\bigintsss_{-\Delta_s/2}^{\Delta_s/2} \bigintsss_{-\infty}^{\infty} \text{d}u'\text{d}u~ e^\frac{-\left(u'-u'_{0,bl}\right)^2}{2\sigma_{u',bl}^2}}
	\label{eqn:slit_trans_eqn}
\end{equation}
where $u'_{0,bl}$ and $\sigma_{u',bl}$ are the central divergence and rms divergence of the beamlet respectively. This can be solved analytically and the general behavior can be seen in the case of $\Delta_s/\Delta_t\gg u'_{0,bl}$ and $\Delta_s/\Delta_t\gg \sigma_{u',bl}$, i.e. the acceptance of the slit is much larger than the divergences in the measured phase space. Then the transmission through the slit can by approximated as
\begin{align}
	T \approx 1 - \frac{\Delta_t}{\Delta_s} &
	\left[ \sqrt{\frac{2}{\pi}}\sigma_{u',bl} e^{-\frac{u^{\prime 2}_{0,bl}}{2\sigma_{u',bl}^2}} \right. \nonumber \\
	&\qquad\left.+u'_{0,bl} \text{erf}\left(\frac{u'_{0,bl}}{\sqrt{2}\sigma_{u',bl}}\right) \right].
	\label{eqn:slit_trans_approx}
\end{align}
The transmission though the slit can be calculated directly from a given simulation output. One example for a simulation is shown in Fig. \ref{fig:slit_acceptance} bottom for a \SI{1}{\nano C} beam measured with a \SI{10}{\micro m} slit. While the result quickly become noisy due to limited number of particles in the beamlets towards the edges of the beam, the general trend matches Eq. \ref{eqn:slit_trans_eqn}. This is despite the non-Gaussian tails in the beam and using Eq. \ref{eqn:beamlet_size} to estimate $\sigma_{u',bl}$. 

The slit transmission can be added to the slit scanner model by removing all macro-particles outside the acceptance of the slit. For standard PITZ measurements using a \SI{1}{mm} thick slit with a \SI{50}{\micro m} opening, the cuts from the slit reduce the emittance by $<0.5\%$. Reducing the slit opening to \SI{10}{\micro m} causes a $<1.5\%$ reduction of the emittance. In both cases, the error can be reduced by scaling up the intensities of each beamlet to account for the losses estimated in Eq. \ref{eqn:slit_trans_approx} despite the assumption of Gaussian which does not hold in the tails. With this correction, the error from the \SI{50}{\micro m} slit is $<0.1\%$ and from the \SI{10}{\micro m} slit is  $<0.5\%$. 

\begin{figure}
	\centering
	\includegraphics{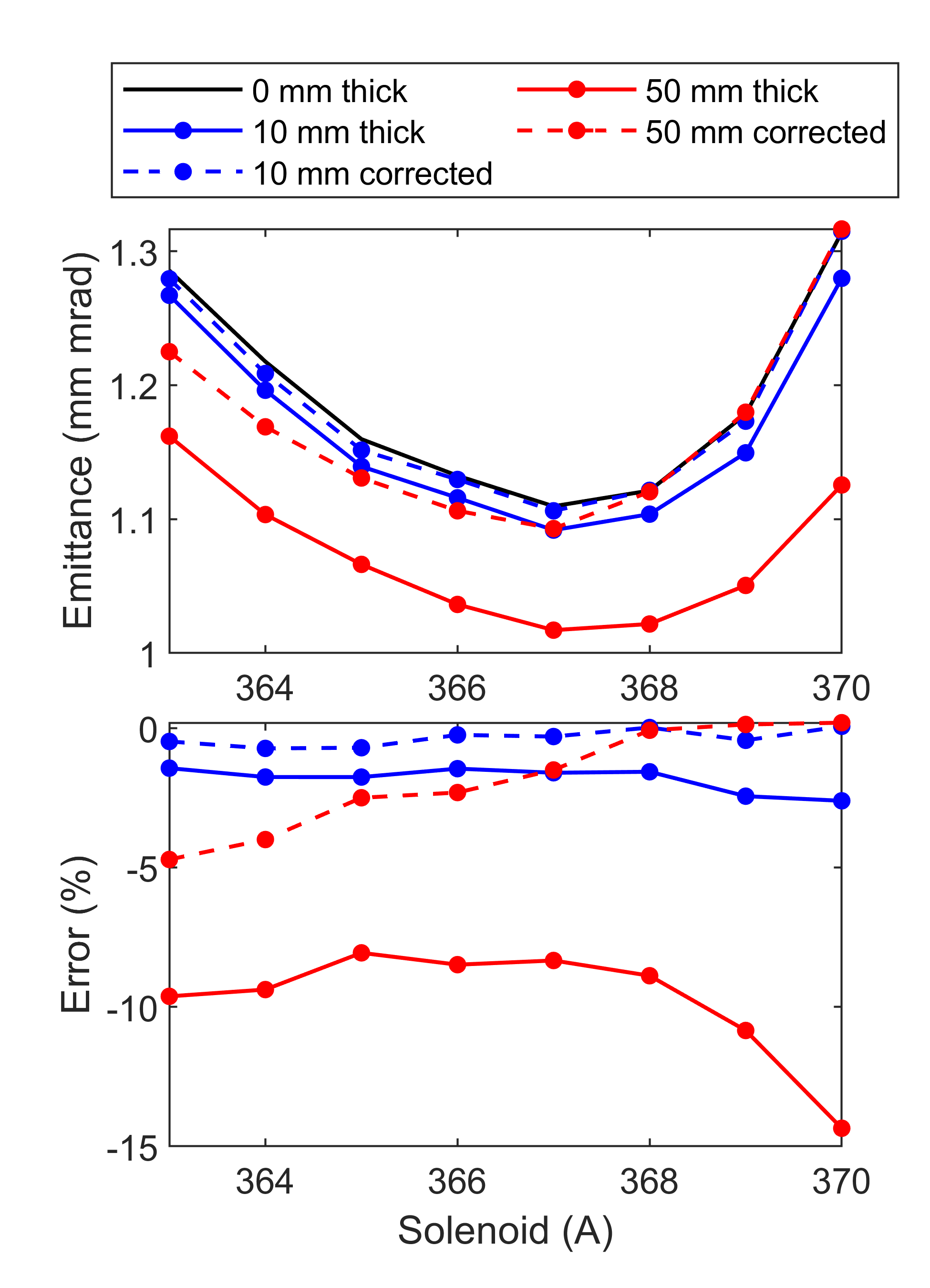}
	\caption{The calculated emittance of a \SI{1}{\nano C} beam using a \SI{10}{\micro m} slit opening and different thickness of the slit. For a \SI{1}{\milli m} slit, the acceptance of the slit can be modeled with Eq. \ref{eqn:slit_trans_approx} and the emittance can be corrected to within 1\%. For \SI{5}{mm} thickness, the acceptance has a more significant effect, and Eq. \ref{eqn:slit_trans_eqn} is no longer a good approximation because larger cuts are being made on non-Gaussian beams.} 
	\label{fig:slit_thickness}
\end{figure}

While this correction is a reasonable estimate when the cuts are small, it can fail. To highlight this, a simulated \SI{1}{\nano C} beam was analyzed with the slit scanner model using a \SI{10}{\micro m} slit opening and slit thicknesses of 0, 1, and \SI{5}{\milli m} (Fig. \ref{fig:slit_thickness}). For the \SI{1}{\milli m} slit thickness, the measured emittance is reduce by up to $\sim2.5\%$ compared to the \SI{0}{\milli m} thickness case. The slit thickness correction reduces the error to within 1\%. When the acceptance of the slit is reduced by increasing the slit thickness to \SI{5}{\milli m}, 5 times thicker than used at PITZ, the measured emittance is reduced by $\sim10\%$ compared to the \SI{0}{\milli m} case. However, Eq. \ref{eqn:slit_trans_eqn} may no longer a good approximation of the transmission because large cuts are being made on phase spaces that are not necessarily Gaussian (see figure \ref{fig:emittance_SC_corr} for example phase spaces). In this case, the correction holds for larger solenoid currents when the phase space is more Guassian, and the error is $<1\%$. But as the solenoid current decreases, the beam deviates from a Gaussian and correction fails resulting is errors $\sim5\%$.

\section{Effect of profiling the beamlets}

\subsection{Imaging resolution}
All profilers have a minimum measurable resolution $\sigma_\text{res}$. This is defined by the slit size for a slit profiler, the wire thickness for a wire scanner or harp, pixel size for imaging with a camera, and multiple scattering in scintillating screens \cite{Spesyvtsev_thesis}. The measured beamlet profiles are a convolution of the true distribution with resolution. This increases the measured beamlet size  
\begin{equation}
	\sigma_{bl,m}^2 = \sigma_{bl,t}^2 + \sigma_\text{res}^2
	\label{eqn:PSF_effect}
\end{equation}
where $\sigma_{bl,t}$ is the true rms size of the beamlet. The measured divergence then becomes
\begin{equation}
	\sigma_{u',m}^2 = \sigma_{u',t}^2 + \frac{\sigma_\text{res}^2}{L^2}.
	\label{eqn:PSF_effect_angle}
\end{equation}

In the case of imaging the beamlets with a scintillating screen and camera, the Point Spread Function (PSF) of the imaging system will cause additional broadening of the beamlet profiles. The measured beam and beamlet images are convolutions of the true distribution, the pixel size $\Delta_p$, and the PSF of the camera/optics system $\sigma_\text{PSF}$. In this case Eq. \ref{eqn:PSF_effect_angle} becomes
\begin{equation}
	\sigma_{u',m}^2 = \sigma_{u',t}^2 + \frac{1}{L^2}\left(\sigma_{PSF}^2 +\frac{\Delta_p^2}{12}\right).
	\label{eqn:PSF_effect_angle_camera}
\end{equation}
The PSF at PITZ is typically $\sim\SI{80}{\micro m}$ rms and increases the measured beamlet size by $\leq\SI{15}{\micro\m}$ for typical beamlet sizes of $\geq\SI{200}{\micro m}$ corresponding to $\sigma_{u',bl}\geq\SI{0.06}{\milli rad}$. Importantly, the stronger the $u-u'$ coupling, the smaller the beamlets will be and, consequently, the impact of the resolution and PSF will be more significant. 

\begin{figure}
	\includegraphics{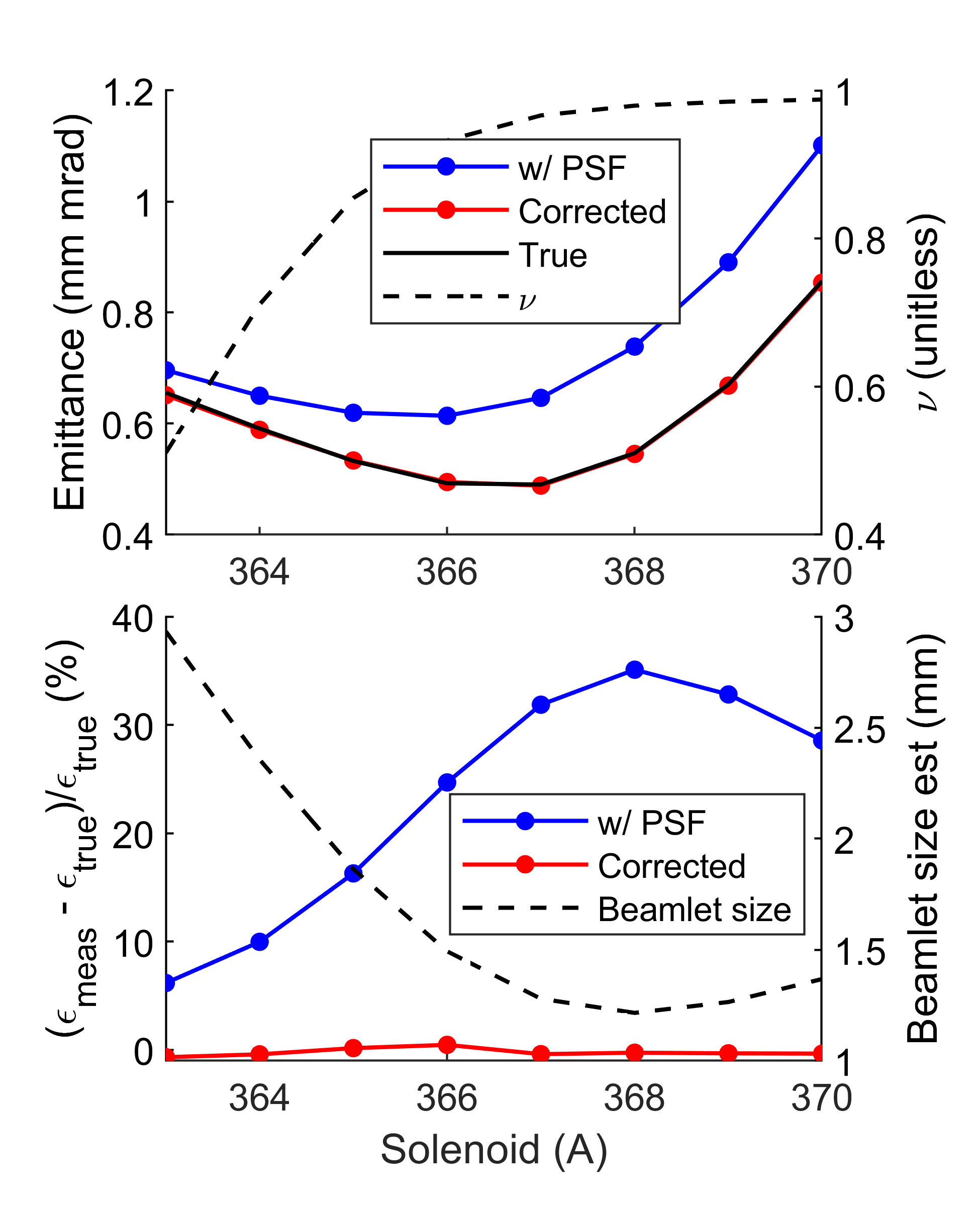}
	\caption{Simulated emittance as a function of gun solenoid current using $\Delta_p=\SI{40}{\micro m}$ and an \SI{80}{\micro m} rms PSF. The measurements match the true values after correcting for the resolution.}
	\label{fig:res_effect}
\end{figure}

To model this effect, the linear slit scan model described above was modified to convolve the created beamlet images with a Gaussian before further processing them to simulate the effect of the imaging resolution. Figure \ref{fig:res_effect} shows the significant impact of the imaging resolution on the measured emittance for PITZ like measurements that are dominated by the PSF with emittance increases $>30\%$. This effect is particularly important when optimizing the gun solenoid current because changing the solenoid current effects the $u-u'$ coupling and the emittance causing the beamlet sizes to vary. The PSF has the most significant impact when the beamlet size is smallest, however the absolute error continues to grow with $\nu$.

\subsection{Profiler step size}
\begin{figure}
	\centering
	\includegraphics{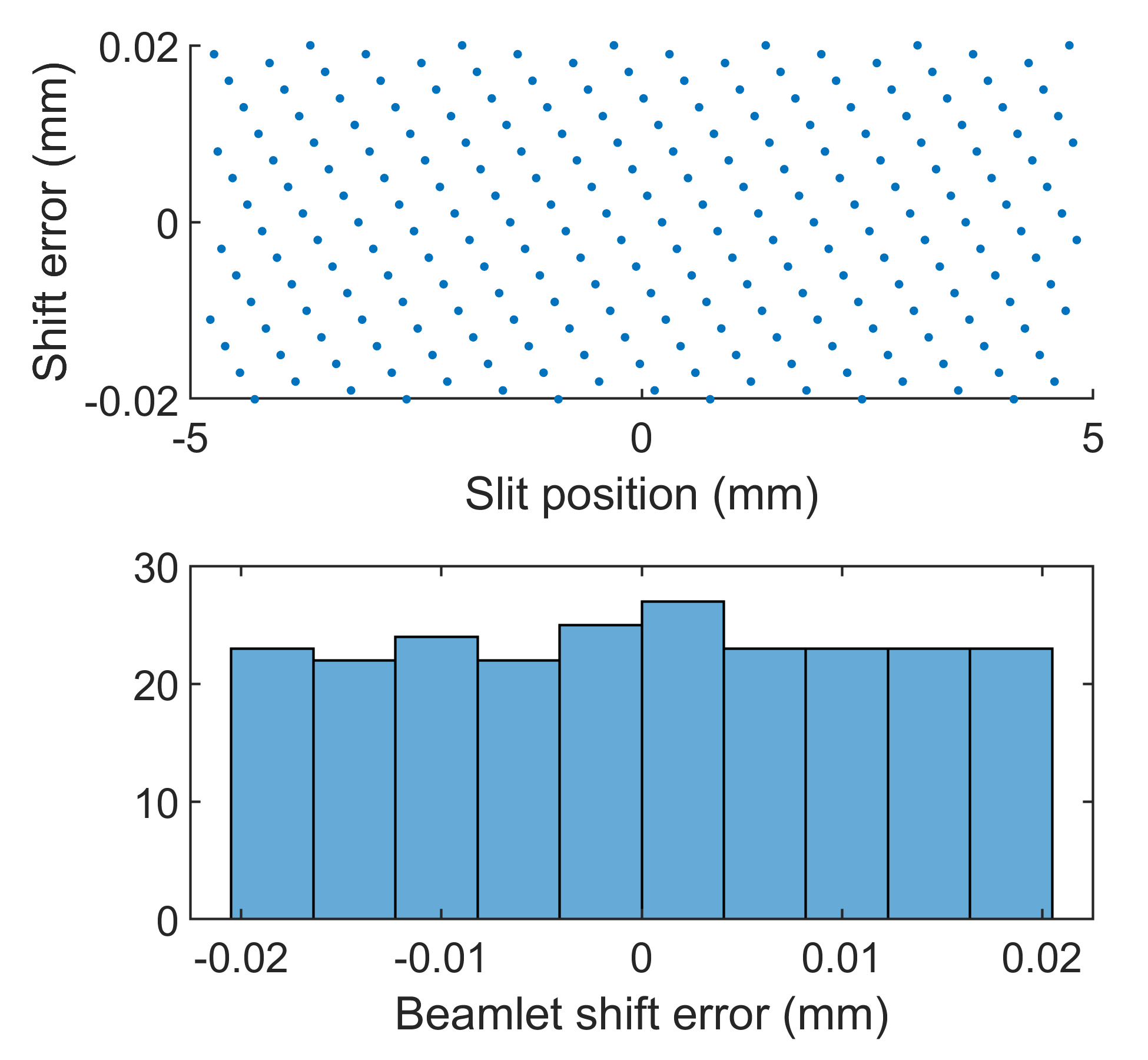}
	\caption{Remainders from shifting the beamlet positions for a slit step of \SI{30}{\micro m} and a profiler step of \SI{41}{\micro m}. The errors are distributed uniformly within the profiler step.}
	\label{fig:shift_error}
\end{figure}

In order to determine the correct divergence distribution of each beamlet, the measured beamlet profile must be shifted by the slit position. This shift, in general, is not an integer multiple of the position step of the profiler $\Delta_\text{step}$, e.g. the wire spacing of a harp. Therefore the shift must be rounded to the nearest $\Delta_\text{step}$ resulting a small offset error and this error is different for each slit position. The errors generally will be uniformly distributed over the profiler step size (Fig. \ref{fig:shift_error}). After the phase space is reconstructed, all the beamlet profiles can be projected to determine the angular profile of the entire beam. This integrates all of the offset errors resulting in an effective convolution of the true distribution with the step size of the profiler further increasing the measured divergence
\begin{equation}
	\sigma_{u',m}^2 = \sigma_{u',t}^2 + \frac{\Delta_\text{step}^2}{12L^2}.
	\label{eqn:pixel_effect}
\end{equation}
The net effect can be made small by taking fine steps with a wire or slit. But for profiles measured with wire harps or camera, the step is fixed to the wire spacing or pixel size respectively. 

For a typical \SI{250}{\pico C} beam emittance measurement at PITZ, the imaged size on a single pixel is \SI{40}{\micro m}, the emittance is $\sim\SI{0.5}{mm~mrad}$, energy is \SI{20}{MeV}, and the beam size is $\sim \SI{0.4}{\milli m}$. This gives a typical rms beamlet size of $\sim\SI{100}{\micro m}$. The \SI{40}{\micro m} pixel has $\lesssim 1\%$ impact on the measured beamlet sizes. However, once again, when the beam is strongly coupled this effect can have a larger impact on the measured emittance resulting in $\sim1\%$ increase in the emittance.

\subsection{Correcting the divergence}
Combining Eqs. \ref{eqn:slit_angle_effect}, \ref{eqn:PSF_effect_angle}, and \ref{eqn:pixel_effect}, the actual measured rms divergence is given by
\begin{equation}
	\sigma_{u',m}^2 = \sigma_{u',t}^2 + \frac{\Delta_s^2}{12L^2} + \frac{\Delta_\text{res}^2}{L^2} + \frac{\Delta_\text{step}^2}{12L^2}. 
	\label{eqn:angle_corr}
\end{equation}
For the case of beamlet measurements with a camera, Eq. \ref{eqn:PSF_effect_angle_camera} should be used for the resolution term and Eq. \ref{eqn:angle_corr} becomes
\begin{equation}
	\sigma_{u',m}^2 = \sigma_{u',t}^2 + \frac{\Delta_s^2}{12L^2} + \frac{\sigma_\text{PSF}^2}{L^2} + \frac{\Delta_{p}^2}{6L^2}. 
	\label{eqn:angle_corr_camera}
\end{equation}
This can be used to correct the measured rms divergence for more accurate calculations of the rms emittance.

If all the parameters in Eq. \ref{eqn:angle_corr} are known, then the measured rms divergence can be corrected using Eq. \ref{eqn:angle_corr} before calculating the emittance. While this method will give the correct emittance it does not correct the measured phase space image. If it is desired to correct the phase space image, then each beamlet image can be deconvolved from all the broadening effects. However, directly deconvolving the beamlet images from the PSF is generally an ill-posed problem and is additionally often not feasible due to the high noise level in images near the edges of the beam. At PITZ, for beamlets near the center, the rms noise level is $\sim$3 orders of magnitude lower than the peak signal. However, near the tails, the signal is close to the noise level. The noise in the images causes significant artifacts when attempting deconvolutions that require strong cuts to remove. This loss of signal significantly alters the measured phase space and should be avoided.

Instead, it is possible to simply estimate the true size of each beamlet using Eq. \ref{eqn:PSF_effect} then scale down the beamlet images to force this rms size while maintaining position of the center of mass of each beamlet. This estimate holds if the beamlets are approximately Gaussian. To test this correction, it was applied to simulated beamlet images used in Fig. \ref{fig:res_effect} and resulted in agreement with the true emittance within 2\% which is a similar level of accuracy to only correcting the rms parameters. However, scaling the beamlet images has little visual impact of the reconstructed phase space and has longer processing time. Therefore it is not used unless detailed analysis of the phase space is desired, such as action-phase analysis \cite{PIP2_allison_scanner, Qian_slit_scan}.

\section{Space charge effects}
\begin{figure}
	\centering
	\includegraphics{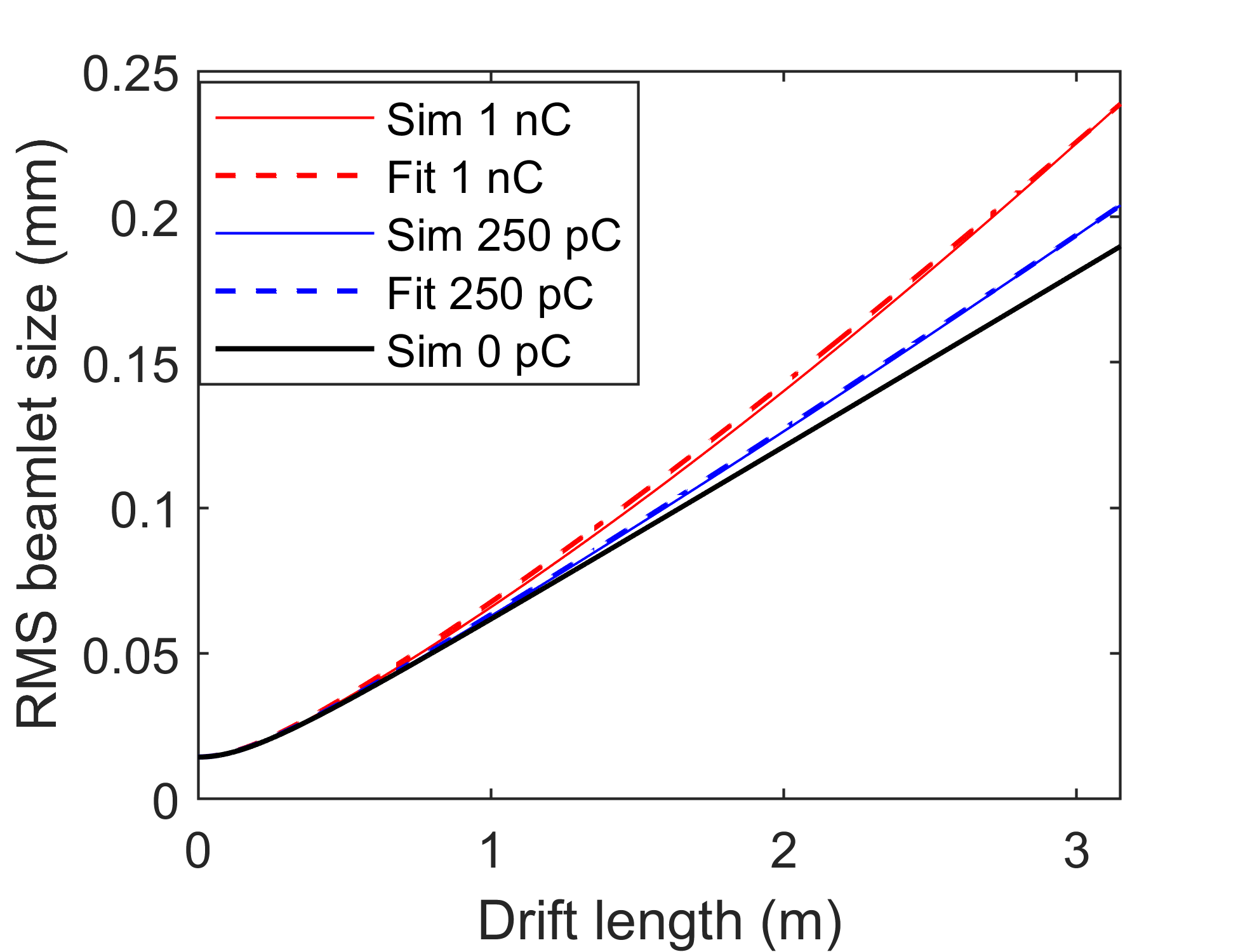}
	\includegraphics{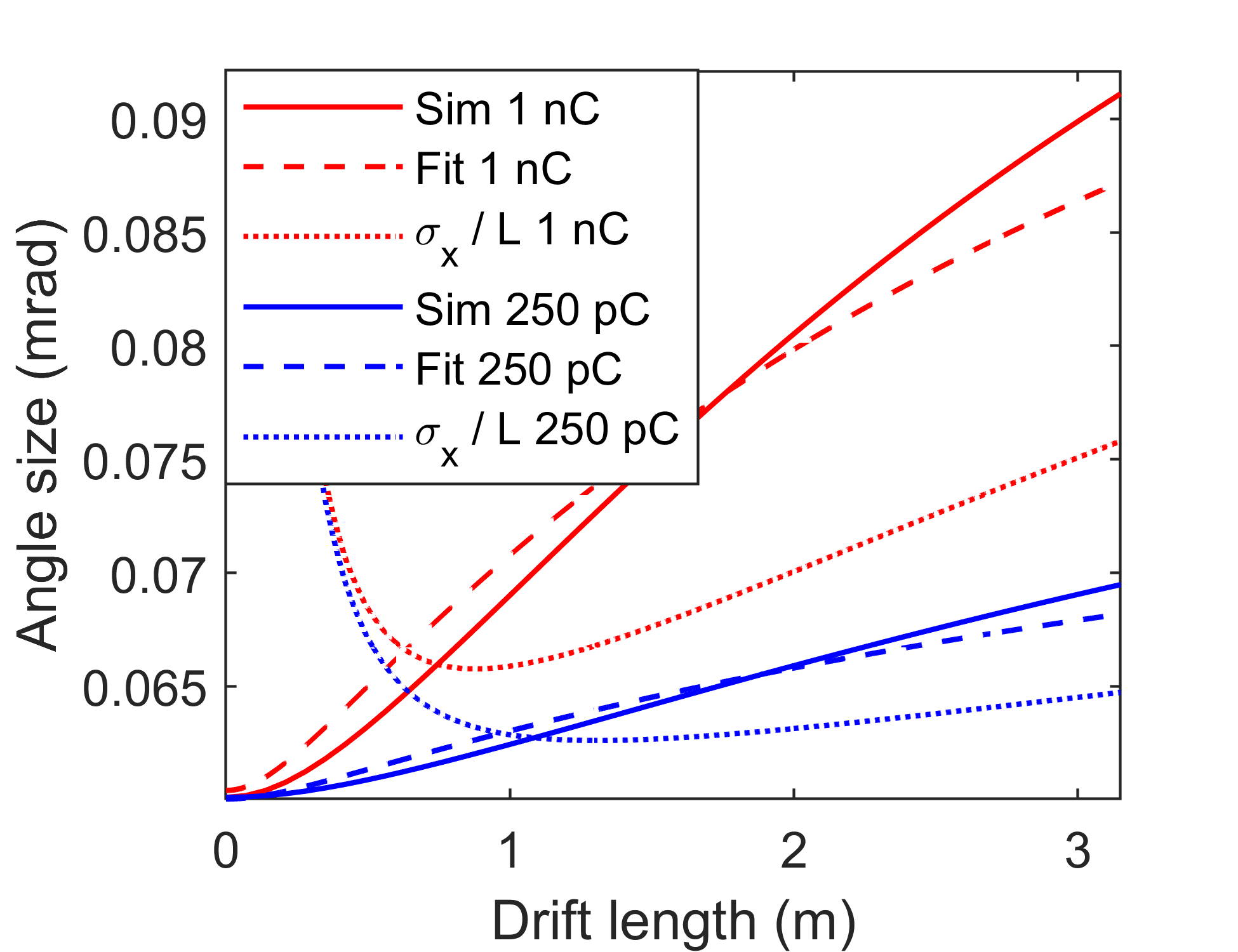}
	\caption{Fitting the central beamlet of a simulated 250 pC beam, slit size = 50$\mu$m. The resulting rms size (top) and rms divergence (bottom) fit well with the true values. The dotted lines are the calculated rms divergence from standard analysis using Eq \ref{eqn:pos_to_angle} which overestimates the rms divergence.}
	\label{fig:fitting_SC}
\end{figure}

\begin{figure*}[t]
	\centering
	\includegraphics{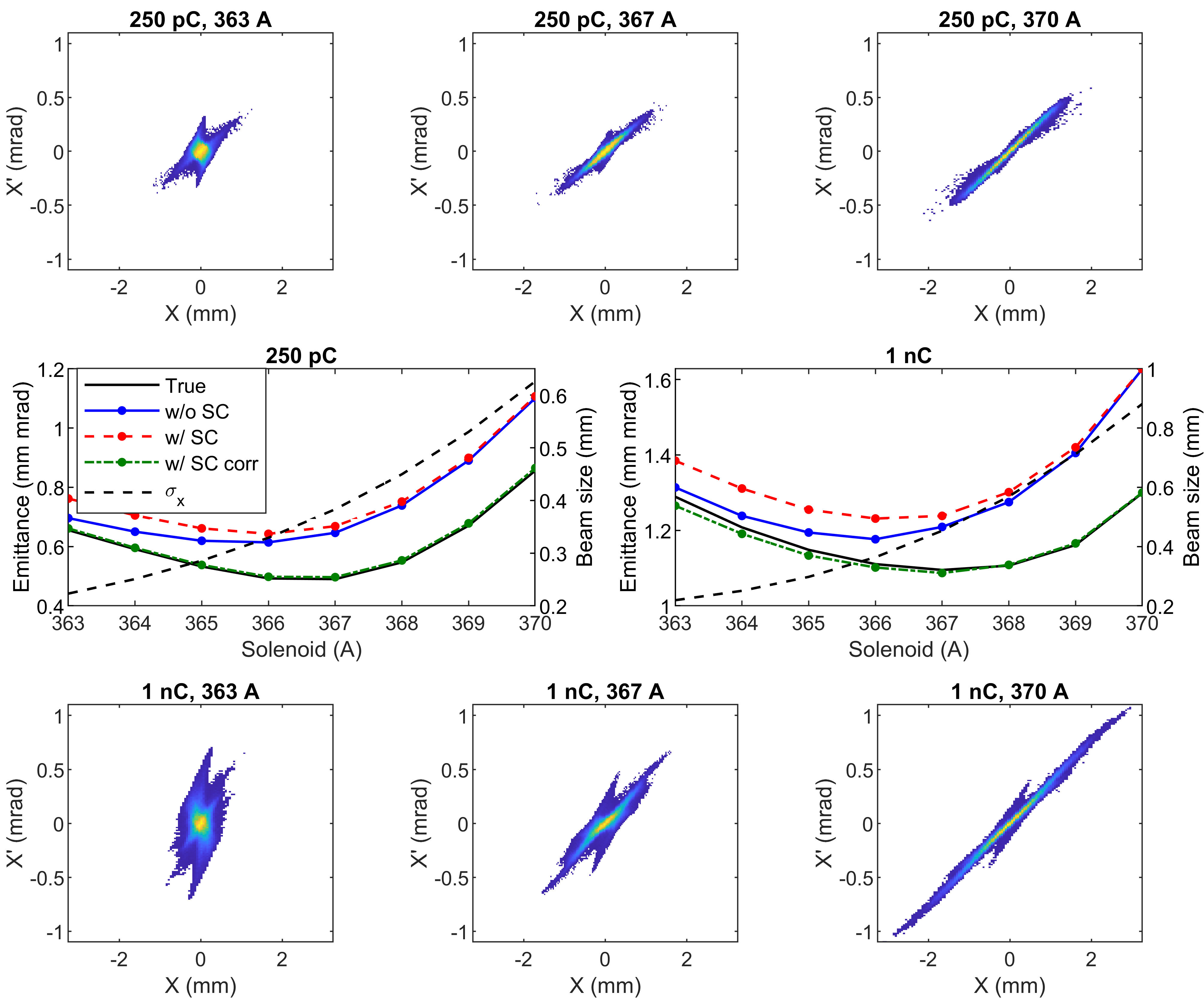}
	\caption{Simulated emittance measurements for a range of gun solenoid currents of \SI{250}{\pico\C} and \SI{1}{\pico\C} beams with bunch lengths of \SI{3.76}{\pico\s} and \SI{5.9}{\pico\s} respectively. The simulations include a \SI{50}{\micro m} slit, \SI{40}{\micro m} pixel size, \SI{80}{\micro m} rms PSF, and space charge effects. After corrections for all the systematic effects the emittances agree with the true values within 2\%. The space charge has the strongest effect when the beam is smallest and the other systematics have a larger impact when the beam is strongly coupled.}
	\label{fig:emittance_SC_corr}
\end{figure*}

The insertion of the slits significantly reduces the passed charge which reduces the effect of space charge during transport of the beamlet to the profiler. However, immediately after the slit, the charge density of the beamlet has not been altered resulting in space charge forces that will increase the beamlet size. From ASTRA simulations of beamlets generated from a \SI{250}{\pico C} and \SI{1}{\nano C} beam using a 50~$\mu$m slit and a drift length of 3.133~m, space charge can cause a $\sim$30\% increase of the beamlet sizes compared to the space charge free case (Fig. \ref{fig:fitting_SC}). Therefore the measured divergence will be overestimated by the same factor. This effect is most prominent in the core of the beam where the space charge density is the highest. The net result is a $\sim$10\% increase in the measured emittance (Fig. \ref{fig:emittance_SC_corr}) with the more significant impact occurring for smaller beams with higher space charge density.

The effect of space charge can be approximately corrected using the rms envelope equation \cite{accel_handbook} without acceleration or focusing and including space charge
\begin{equation}
	\sigma_x'' = \frac{\epsilon_{x,g}^2}{\sigma_x^3} + \frac{I}{I_A(\sigma_x + \sigma_y)\gamma^3}
	\label{eqn:rms_model}
\end{equation}
where the slit is assumed by cut in the $x$ plane, $\gamma$ is the Lorentz factor, $I$ is the beam current, and $I_A$ is the Alfven current. $I$ is defined by $Q_\text{beamlet}/3\sigma_T$ where $Q_\text{beamlet}$ is the charge of the beamlet, $\sigma_T$ is the rms temporal bunch length, and the factor of three was chosen to best match the analytic model to simulation results. This model can be used to determine the rms divergence of the beamlet at the slit by varying the initial rms divergence $\sigma_{x',0}$ and fitting to the final beamlet size on the screen $\sigma_{x_f}$. The initial conditions are the initial beamlet size, given by the slit size, and the assumption that there is no $x-x'$ coupling at the slit. The initial emittance of the beamlet in the cut plane is defined at the slit to be the product of the rms slit size and the initial rms divergence. The rms parameters of the uncut plane are also needed because the equations of motion are coupled. The model is sensitive to the beam size in the uncut plane as this directly affects the space charge density at the slit. This is measured at PITZ by the spot image for every emittance scan, or an emittance scan in the other plane can be used. The other second order moments in the uncut plane do not significantly impact the model unless they are significantly off, $\gtrsim20\%$, and the results from analysis without accounting for space charge can be used.  In addition, the beam current is needed, which is determined from the measured beam charge and temporal profile which can be measured with a transverse deflecting structure \cite{PITZ_TDS}.

The fitting is performed with the rms divergence as a free parameter and using a Runge-Kutta solver to propagate the beamlet parameters to the profiler and fitting the beamlet size to the measured value after correction with Eq. \ref{eqn:angle_corr}. The optimized beamlet size agrees well with beamlet simulated with ASTRA (Fig. \ref{fig:fitting_SC}). There is some discrepancy in the beamlet divergence due to the model assuming a Gaussian beam in all dimensions, but the beamlet density will be nearly uniform across the slit. Once the corrected rms divergence $\sigma_{u',bl,\text{fit}}$ is found for a beamlet, the beamlet image can then be corrected in the same manner as deconvolving the PSF. The beamlets are scaled such that 
\begin{equation}
	\sigma_{u,\text{beamlet}} = \sigma_{u',bl,\text{fit}} L.
	\label{eqn:angle_scaling}
\end{equation}
The phase space can then be reconstructed using Eq. \ref{eqn:pos_to_angle} which doesn't account for space charge.

This correction was tested with ASTRA simulations that propagated beamlets with space charge from the slit to the profiler for all beamlets in a slit scan measurement. Each simulated beamlet images was then corrected using the fitting described above. After correction, the measured emittance is within 2\% of the true emittance for 250~pC and 1 nC results (Fig. \ref{fig:emittance_SC_corr}) despite the errors from non-Gaussian beamlets.

\section{Noise cuts}
\begin{figure}
	\centering
	\includegraphics{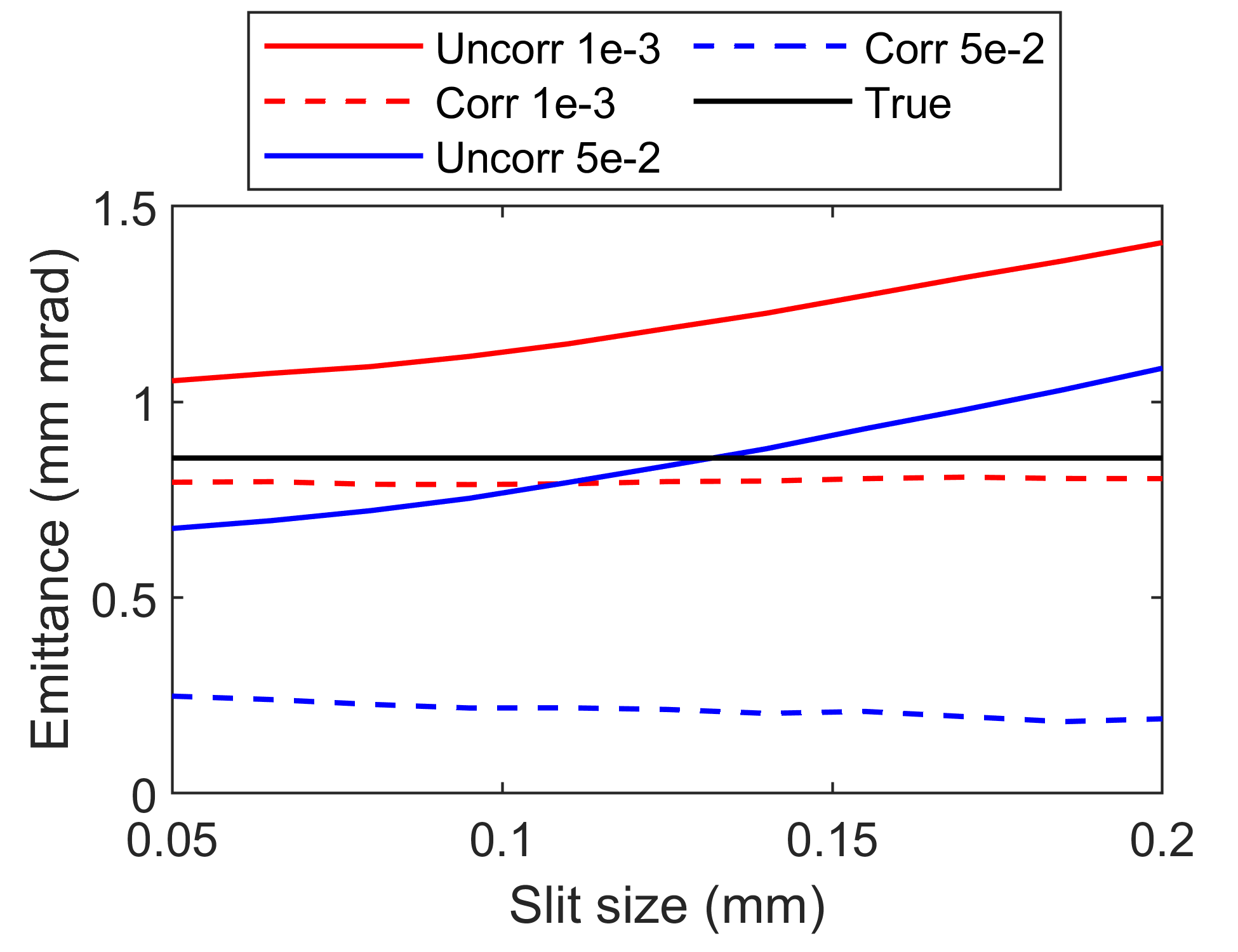}
	\caption{The measured and corrected emittance for simulations at \SI{370}{A} using a range of slit sizes to generate the beamlets. The beamlet images are cut at $0.001$ and $0.05$ with respect to the peak intensity. 
	The measured emittance after corrections is relatively flat compared to the initial measured values. This holds even with the large cut of $0.05$ which no longer gives a reasonable measurement of the emittance. Similar trends are seen when scanning other parameters at with different beam distributions.}
	\label{fig:cut_scaling}
\end{figure}

Up to this point, the studies have assumed no signal losses in the measurements. However, in reality, the beamlet measurements are noisy and all information below the noise floor is lost. Most of the corrections described above rely on calculated rms parameters, which will be reduced when a noise cut is made. Therefore it is necessary to ensure the feasibility of the corrections in the presence of cuts. To test this, one of the scan corrections parameters, e.g. the slit size, was scanned in the linear slit scanner model and for each setting noise cuts were made to the generated beamlet images. The stability of the calculated parameters could then be checked before and after corrections for different cut levels. For these studies it is required to convolve the beamlet images with a PSF to  artificially reduce the discretization caused by the finite number of particles in the simulation. The noise cuts are made at the same absolute intensity level for all beamlet images. That level is defined by a fraction of the peak signal across all beamlets.  The results of corrections with different slit width are shown in Fig. \ref{fig:cut_scaling} using the simulated distribution at \SI{370}{A} because this distribution is most sensitive to the slit size. With a cut level of $10^{-3}$, similar to the cut level of measurements at PITZ, the corrected emittance is mostly constant even up to very large slits of \SI{200}{\micro m}. For a cut level of $0.05$ the corrected emittance has more variation, but is still fairly flat considering that the corrected emittance is no longer a reasonable measure of the true emittance. This study was repeated for different beam distributions and scanning different scanner parameters with similar results. This consistency of the corrected emittance means the corrections are reliable even for fairly large cuts.

\begin{figure}
	\centering
	\includegraphics{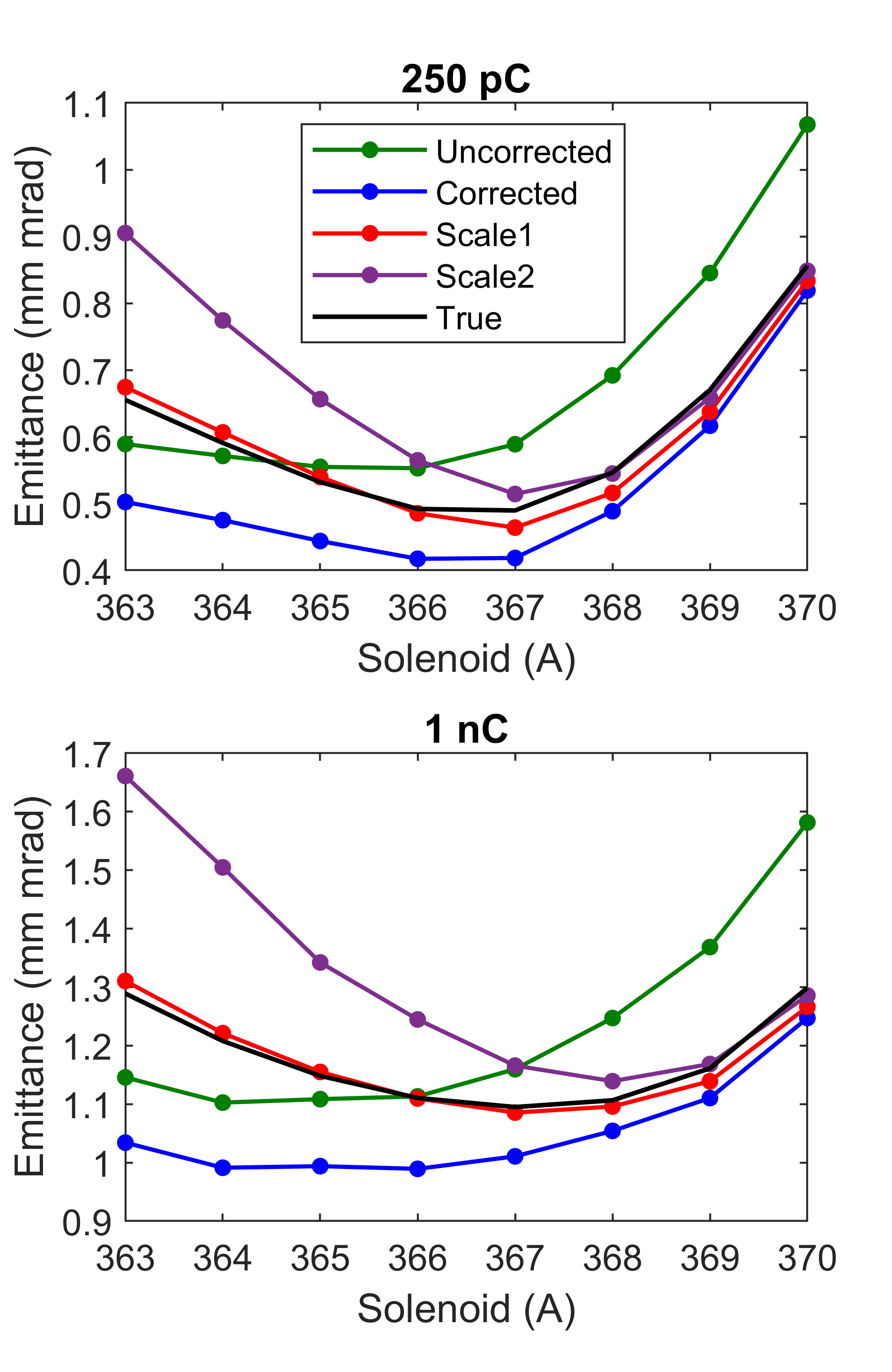}
	\caption{The simulated emittance measurements from Fig. \ref{fig:emittance_SC_corr} with the addition of a  cut level of $10^{-3}$ applied to the beamlet images. The raw measurements do not show a clear minimum emittance. After corrections are applied, the trend of the emittance is improved, but there remains a significant discrepancy with the true emittances. The \textit{scale1} emittance (Eq. \ref{eqn:scale_factor}) gives good agreement with the true emittances while the \textit{scale2} emittance significantly overestimates the true emittance.}
	\label{fig:sol_scan_cuts}
\end{figure}

Even after corrections, the effect of the noise cuts on the calculated beam parameters is sensitive to the cut level and the nature of the beam tails. The core-tail distribution and noise level can be different between different measurements which makes directly comparing measurements with noise cuts less straight forward. For example, when scanning the strength of the solenoid at the photo gun to optimize the emittance compensation scheme alters the nature of the tails as they rotate in phase space and manifest differently in the beamlets. The noise cuts therefore cut different parts of the beam and, after cuts, the setting for optimal emittance can be distorted or not present. For example, Fig. \ref{fig:sol_scan_cuts} shows the simulated emittance for a gun solenoid scan with cut levels on the beamlet images of $10^{-3}$. Due to the nature of the core-tail distributions, the emittance measured with lower solenoid currents see the most significant impact from the cut resulting in a mostly flat trend that obscures the minimum emittance. In addition, the difference in the effect of these cuts, in conjunction with the systematic effects discussed above, result is cases where the measured emittance can be smaller or larger than the true emittance depending on the beam distribution. However, after applying the corrections for residual space charge and systematic errors, the minimum can once again be found, and the emittance is always less than the true value. However the behavior as a function of the solenoid current still differs from the true emittance.

In the presence of noise cuts on the beamlet images, it was shown that the 100\% emittance can be estimated by \cite{Qian_slit_scan}
\begin{equation}
	\epsilon_{100} \approx \epsilon_{s2} \approx \frac{\sigma_{u,true}^2}{\sigma_{u,scan}^2}\epsilon_{m}
\end{equation} 
where $\epsilon_{s2}$ is the '\textit{scale2}' emittance, $\sigma_{u,true}$ is the true rms beam size, $\sigma_{u, scan}$ is the rms size from the slit scan, and $\epsilon_m$ is the measured emittance. At PITZ, $\sigma_{u,true}$ is estimated by directly measuring the beam profile at the location of the slit with a YAG screen. This image will have much better SNR than the low intensity beamlet images and, therefore, is a reasonable estimate of $\sigma_{u,true}$. However, while this correction factor, referred to here the \textit{scale2} factor, gives accurate results for Gaussian beams, for realistic beams, it was empirically found that the \textit{scale2} factor tends to be overly sensitive to the tail distribution. This can result in a wide range of behaviors from dramatically overestimating the emittance to barely changing the measured emittance (Fig. \ref{fig:sol_scan_cuts}). 

Instead, a more conservative estimate can be used \cite{PITZ_overview}
\begin{equation}
	\epsilon_{s1} \approx \frac{\sigma_{u,spot}}{\sigma_{u,scan}}\epsilon_{m}
	\label{eqn:scale_factor}
\end{equation} 
This '\textit{scale1}' emittance is still sensitive to the exact nature of the cut tail distribution, but it is not prone to large changes in the emittance. From simulations with realistic beam distributions, the \textit{scale1} emittance is generally near true emittance for a much larger range of solenoid currents than the \textit{scale2} emittance. Therefore, the \textit{scale1} emittance can help with gun solenoid scans to reconstruct the basic trend near the minimum emittance. In Fig. \ref{fig:sol_scan_cuts}, after this scale factor is applied to the corrected emittances, the reduction of the emittance for smaller beams is improved. This results in emittance behavior that more closely follows the true behavior. 

\section{Measurements}
\begin{figure*}
	\centering
	\includegraphics{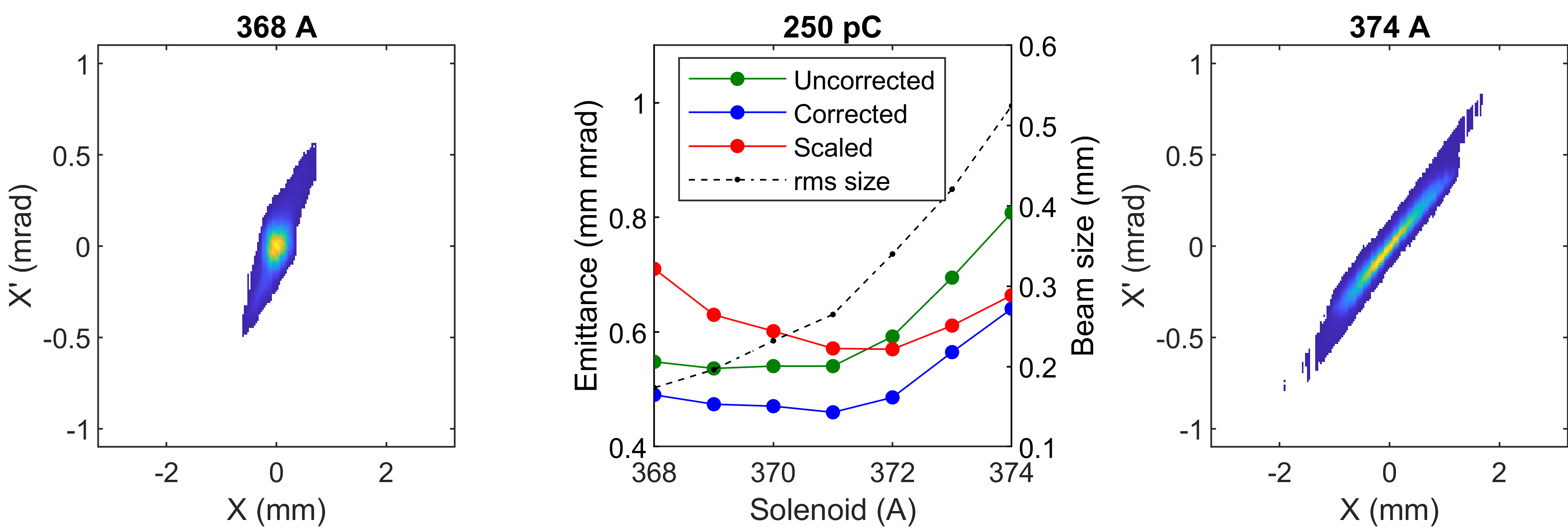}
	\caption{Measured emittances at PITZ of a \SI{250}{pC}, \SI{17}{MeV/c} beam with a \SI{50}{\micro m} slit. The initial measurements have a nearly constant emittance for smaller beam sizes; similar to simulations with cuts (Fig. \ref{fig:sol_scan_cuts}). After corrections and scaling, a clear minimum is found at \SI{372}{A}.}
	\label{fig:measured_emittance}
\end{figure*}

These corrections were applied to emittance measurements at PITZ for a \SI{17}{MeV/c}, \SI{250}{\pico C} beam. Shown in Fig. \ref{fig:measured_emittance} are the measured emittance at PITZ for a gun solenoid scan using a \SI{50}{\micro m} slit. Similar to simulations in Fig. \ref{fig:sol_scan_cuts}, the uncorrected results show fairly flat behavior at lower solenoid currents making determining an optimal setting difficult. The minimum emittance occurs at \SI{369}{A} but its value is similar for 368-\SI{371}{A}. After applying the corrections for slit size, imaging, and space charge, the resulting emittances are reduced by 10-20\% and a minimum becomes slightly more clear. Then after the addition of the \textit{scale1} factor to account for noise cuts, there is a much more clear minimum emittance at \SI{372}{A} which is expected to be closer to the true optimal emittance. This shift in the optimal solenoid current after the corrections causes a significant change in the rms beam size at the scanner, going from \SI{0.2}{mm} at \SI{369}{A} to \SI{0.34}{mm} at \SI{372}{A} which can significantly change the tuning of the rest of the beamline.

\section{Summary}
The systematic errors in slit-profiler measurements can have significant impact on the measured emittance even when efforts are made to minimize their effect. The size of the slit causes a small increase on the measured rms beam size and divergence and, in addition, the slit opening imprints a $u\text-u'$ coupling to the measured phase space image. These errors are generally negligible for each rms moment, however when $\sigma_{uu}/\sigma_u\sigma_{u'}\approx1$ then these small errors can cause a significant increase in the calculated emittance. Other effects, including the profiler resolution, PSF, and step size also cause broadening of the beamlet profiles resulting in a larger emittance. These effects can be easily accounted for by correcting the measured rms parameters by the rms size of the effects (Eqs. \ref{eqn:slit_profile_effect}-\ref{eqn:slit_coupling_effect}, \ref{eqn:angle_corr}).

Residual space charge forces in the beamlets can also increase the measured emittance. The impact of space charge is most prominent for smaller beams and higher charges where the space charge density is higher. This effect can be compensated using a numeric transport model to determine the true divergence of the beamlets by fitting the measured beamlet sizes.

Lastly, the effects of noise cuts were studied. The effect of the noise cuts is the most unpredictable because it depends strongly on the beam core and tail/halo distributions and how they manifest themselves in the beamlets. This makes comparing measured phase spaces challenging because different parts of the beam are being cut. While exact corrections of the emittance cannot be made for cuts, an approximation can be made by scaling the emittance with Eq. \ref{eqn:scale_factor}.

When all these corrections are applied, then the resulting emittance can be reduced by >\SI{0.1}{mm~mrad} and the emittance behavior when scanning a parameter can change. In particular, the increase of the measured emittance from systematic effects is larger when the beam is more strongly $u\text-u'$ coupled and the residual space charge has the largest impact when the beam size is smallest. But correcting for these effects, even in the presence of noise cuts, results in good agreement with the true emittance.

\section{Acknowledgments}
This work was supported by EURIZON and the European XFEL research and development program.

\bibliography{Emittance_correction.bib}
\bibliographystyle{unsrt}

\end{document}